\documentclass{mem}

\usepackage{txfonts}
\usepackage{balance}
\usepackage{graphicx}
\usepackage[a4paper,breaklinks,dvipdfm]{hyperref}
\usepackage{natbib}
\bibpunct{(}{)}{;}{a}{}{,}

\idline{83,~n.3}{1}
 
\include{amssymbol}
\input{epsf}

\def\to{\rightarrow}  
  
\def\mpc{\mbox{Mpc}}

\def\AJ{{\it Ap. J.} }

\def\AJS{{\it Ap. J. Supp.} }

\def\CQG{{\it Class. Quantum Gravity} }

\def\GRG{{\it Gen. Relativity and Gravitation} }

\def\IJMP{{\it Int. J. Mod. Phys.} }

\def\MNRAS{{\it Mon. Not. R. Ast. Soc.} }

\def\PL{{\it Phys. Lett.} }
\def\PR{{\it Phys. Rev.} }

\def\al{\alpha} \def\be{\beta}  
   
\def\th{\theta}   \def\ka{\kappa}
   
\def\si{\sigma}   
   
\def\om{\omega}  \def\De{\Delta} 
\def\La{\Lambda}   
 \def\Om{\Omega} \def\mn{{\mu\nu}} \def\cl{{\cal L}}

 \def\frac#1#2{{\textstyle{{#1}\over
{#2}}}} 
\def\lsim{\mathrel{\rlap{\lower4pt\hbox{\hskip1pt$\sim$}}
\raise1pt\hbox{$<$}}}
\def\gsim{\mathrel{\rlap{\lower4pt\hbox{\hskip1pt$\sim$}}
\raise1pt\hbox{$>$}}} \def\sqr#1#2{{\vcenter{\vbox{\hrule height.#2pt
\hbox{\vrule width.#2pt height#1pt \kern#1pt \vrule width.#2pt} \hrule
height.#2pt}}}}
\def\square{\mathchoice\sqr66\sqr66\sqr{2.1}3\sqr{1.5}3}
\def\beq{\begin{equation}} \def\eeq{\end{equation}}
\def\beqa{\begin{eqnarray}} \def\eeqa{\end{eqnarray}}
\def\eq#1{Eq. (\ref{#1})}

\begin{document}

\title{Testing a non-minimal coupling\\between matter and curvature}

\author{
J. \,P\'aramos
          }

\institute{
Instituto de Plasmas e Fus\~ao Nuclear, Instituto Superior T\'ecnico\\Avenida Rovisco Pais 1, 1049-001 Lisboa, Portugal \\
\email{paramos@ist.edu}
}

\authorrunning{J. P\'aramos}

\titlerunning{Testing a non-minimal coupling between matter and curvature}

\abstract{
One of the most interesting and current phenomenological extensions of General Relativity is the so-called $f(R)$ class of theories; a natural generalization of this includes an explicit non-minimal coupling between matter and curvature. The purpose of this work is to present a unified view of the applicability of the latter to various contexts, ranging from astrophysical matter distributions to a cosmological setting. Various results are discussed, including the impact of this non-minimal coupling in the choice of Lagrangian density, a mechanism to mimic galactic dark matter and a Cosmological Constant at a astrophysical scale, the possibility of accounting for the accelerated expansion of the Universe and modifications to post-inflationary reheating. The equivalence between a model exhibiting a non-minimal coupling and multi-scalar-theories is also discussed.
}

\maketitle

\section{Introduction}

Cosmology currently faces two outstanding challenges: the origin of dark energy and dark matter. Research on these dark components of the Universe may be roughly divided into two fields: one assumes that Einstein's General Relativity (GR) is valid at the relevant energy scales, thus focusing on the search for the missing matter or energy contribution; the other posits instead that dark matter and energy merely reflect deviations of gravity from GR.

When addressing dark matter, the first approach relies fundamentally on the characterization of additional matter species, arising mostly from extensions to the Standard Model of particles and fields, collectively dubbed as weak-interacting massive particles (WIMPS): these include {\it e.g.} neutralinos, which arise naturally from supersymmetry, or the axion, derived from an elegant solution to the strong CP problem \citep{Feng}.

Similarly, the nature of dark energy usually involves the addition of a new scalar field, which slow-rolls down an adequate potential ---   quintessence \citep{Copeland}. A putative unification of dark matter and energy has also been suggested, by resorting to a scalar field \citep{Rosenfeld} or an exotic equation of state (EOS), such as the generalized Chaplygin gas \citep{Chaplygin1,Chaplygin2,Chaplygin3}.

The second approach follows a quite different route, by assuming that GR itself is at fault: instead of dark matter or energy, one may consider that dark gravity is at play. Thus, several attempts to generalize GR have surfaced: these are usually phenomenological in nature, positing putative low-energy modifications of Einstein's theory that should be derived from a yet unknown high-energy, fundamental theory of gravitation. In a cosmological context, modifications of the Friedmann equation to include higher order terms in the energy density $\rho$  have been proposed \citep{Maartens,cardassian,scalar}, as well as considerations of the impact of a van der Waals EOS for matter \citep{vdW}.

Another widely discussed candidate is the MOdified Newtonian Dynamics (MOND), a non-relativistic modification of the Poisson equation, which becomes $\nabla [\mu(\nabla \Phi/a_0) \nabla \Phi] = 4\pi G \rho$, where $\Phi$ is the gravitational potential, $\rho$ the density and $\mu(x)$ a function satisfying $\mu(x) \sim x $ for $x\ll 1$ and $\mu(x) \sim 1$ for $x \gg 1$. By introducing the typical acceleration $a_0 \sim 10^{-10}~m/s^2$, MOND allows a simple derivation of the flattening of the rotation curves of galaxies, thus accounting for the missing ``dark'' matter challenge \citep{MOND}.

However, MOND has been shown to suffer from one outstanding issue, namely the inability to explain the gravitational profile of the so-called Bullet cluster \citep{bullet} --- although a proposal resorting to heavy neutrinos with a mass close to $\sim 2~{\rm eV}$ has been put forward \citep{MONDbullet1,MONDbullet2}. On the theoretical side, the current relativistic formulation of MOND, dubbed Tensorial-Vector-Scalar (TeVeS) theory, resorts to two scalar (one dynamic and one auxiliary) and one vector field, together with a physical metric $\tilde{g}_\mn$ used to couple with matter \citep{TeVeS}: this formulation may scarcely be considered an elegant one, and one is tempted to call these extra fields as dark matter in another guise.

Changes to the action functional provide a simple pathway for implementing extensions to GR: a straightforward approach replaces the linear scalar curvature $R$ term in the Einstein-Hilbert action with a function $f(R)$ \citep{fR1,fR2} (see \citep{felice} for a review); more evolved dependences may also be explored, as considered in Gauss-bonnet models \citep{GB}.

This class of $f(R)$ theories enjoys considerable success in several fronts, including the puzzle of the missing ``dark'' matter in galaxies and clusters \citep{DMfR,ClustersfR}, as well as the nature of ``dark'' energy \citep{capoexp}; the early period of rapid expansion of the universe is described by the Starobinsky inflationary model $f(R)=R + \al R^2$ \citep{Staro}; local Solar system impact and the related parameterized post-Newtonian (PPN) metric formalism have also been studied  \citep{PPN}.

The mathematical equivalence between $f(R)$ and scalar-tensor theories has proven very profitable, given the possibility of comparing results between the two types of model \citep{felice}. Further insight is gained from considering the so-called Palatini formulation, where both the metric and the affine connection are taken as independent variables \citep{Palatini} --- an alternate to the usual metric affine connection approach, where the affine connection is taken {\it a priori} as depending on the metric.

Given the proficuous results arising from $f(R)$ theories, one is naturally tempted to further generalize this model --- in the process extending its explanatory capability. Thus, another interesting possibility has been subject to scrutiny: not only that the curvature is non-trivial in the Einstein-Hilbert Lagrangian, but also that the coupling between matter and geometry is non-minimal \citep{Lobo}, as written below:
\beq S = \int \left[ \ka R + f_2(R) \mathcal{L} \right] \sqrt{-g} d^4 x~~. \label{model}\eeq
\noindent where $\ka = c^4/16\pi G$.

The purpose of this work is to present an unified view of several results that arise from this framework in a wide range of scenarios \citep{analogyf2,mimic,localCC,fluid,accexp,preheatingf2} ; these were developed in collaboration with O. Bertolami, F. S. N. Lobo and P. Fraz\~ao.

\section{The model}

Variation with respect to the action \eq{model} yields the modified Einstein field equations,
\beqa \label{field0} && 2\left(\ka F_1+  F_2 \cl \right) G_\mn =  2 \De_\mn \left(\ka F_1 + F_2 \cl\right) \\ \nonumber && - [ \ka( F_1R - f_1 ) + F_2 \cl R] g_\mn + f_2 T_\mn ~~, \eeqa
\noindent where one defines $\De_\mn \equiv \nabla_\mu \nabla_\nu - g_\mn \square$ and $F_i(R) \equiv f'_i(R)$. As expected, GR is recovered by setting $f_1(R) = R $ and $f_2(R) = 1$.
By taking the trace of the above, one gets
\beqa \label{fieldtrace} && \ka (F_1 R - 2f_1) +  F_2 \cl  R =  \\ \nonumber &&  {1 \over 2} f_2 T - 3\square \left(\ka F_1 + F_2 \cl\right) ~~, \eeqa
\noindent where $T$ is the trace of the energy-momentum tensor. The usual $f(R)$ theories are recovered by setting $f_2(R) = 1$, so that \eq{fieldtrace} yields an algebraic relation for $R=R(T)$. However, if one considers a non-minimal coupling $f_2(R) \neq 1$, the above becomes a differential equation: in particular, this enables the possibility that, given a rapidly varying $F_2 \cl$, the above may lead to high curvatures even if $T$ is low.

\subsection{Non-conservation of energy}

The Bianchi identities may be used to derive the non-(covariant) conservation of the energy-momentum tensor,
\beq \nabla_\mu T^\mn={F_2 \over f_2}\left(g^\mn \cl-T^\mn\right)\nabla_\mu R~~. \label{cov} \eeq
\noindent In the absence of a non-minimal coupling, $f_2(R)=1$, one recovers the covariant conservation of the energy-momentum tensor. 

\eq{cov} may be rewritten as an extra force imparted on test particles, so that its trajectory will deviate from geodesical motion. Thus, the Equivalence Principle may be broken if the {\it r.h.s.} of the last equation varies significantly for different matter distributions.

In Jordan-Brans-Dicke theories, the curvature appears coupled to a scalar field \citep{Damour}, leading to expressions similar to \eq{cov}, due to the energy exchange between matter and the latter. However, this non-conservation may be transformed away by a suitable conformal transformation to the Einstein frame, where the curvature appears uncoupled \citep{conformal}. By the same token, one could assume that the non-conservation law \eq{cov} is only an artifact of the particular frame adopted, not a physically significant feature of the model.

However, performing the conformal transformation $ g_\mn \rightarrow \tilde{g}_\mn = f_2 g_\mn $, one finds that energy conservation is attained, $ \tilde{\nabla}_\mu\tilde{T}^\mn = 0 $, only if $\tilde{T}^\mn = f_2^{-2} T^\mn$ and $ 2\cl = T $. Unless this relation holds for all matter species, energy conservation is indeed broken.

Thus, it suffices to use one example that breaks this relation: taking the case of a perfect fluid, with a Lagrangian density $\cl = -\rho$ \citep{fluid}, then one obtains $p = -\rho/3$ --- signaling a perfect fluid with negative pressure (but different from a cosmological constant,where one has $p_\La = -\rho_\La$); if one instead opts for the Lagrangian density $\cl = p$ \citep{Sotiriou1}, then $p = \rho$ is obtained --- the EOS for ultra-stiff matter.

The discussion of which form for $\cl$ is correct in the presence of a non-minimal coupling is addressed in the following section; nevertheless, it is clear that one cannot recover energy conservation for a perfect fluid --- and, by extension, for all matter species: this non-conservation law is indeed a fundamental property of the model under scrutiny.

\section{Equivalence with multi-scalar-tensor models}

Through a suitable conformal transformation, the usual $f(R)$ theories can be rewritten as GR with an added scalar field contribution, which is dynamically identified with the curvature, $\phi = R $. Similarly, the discussed non-minimally coupled model \eq{model} can be recast as a multi-scalar field theory, with two scalar fields, albeit a ``physical'' metric remains in the matter Lagrangian \citep{analogyf2}. Indeed, by performing a conformal transformation $g_\mn \rightarrow \tilde{g}_\mn = \exp[(2/\sqrt{3})\varphi^1] g_\mn$, the equivalent action is obtained,
\beqa  S & = & \int  \sqrt{-\tilde{g}} d^4x \bigg[ f_2(\varphi^2)  e^{-{2 \over \sqrt{3}}\varphi^1} \cl(g_\mn,\chi) \\ \nonumber && + 2 \ka \left( R - 2\tilde{g}^\mn \si_{ij} \varphi^i_{,\mu} \varphi^j_{,\nu} -  4 
U \right) \bigg]~~, \eeqa
\noindent where $\chi$ denotes all matter fields, $\varphi^1$ and $\varphi^2$ are scalar fields related to the scalar curvature and matter Lagrangian density through
\beq \varphi^1 = {\sqrt{3}\over2} \log \left[ {F_1(R) + F_2(R) {\cal L} \over 2\ka} \right] ~~, \eeq
\noindent and $\varphi^2 = R $; $\si_{ij}$ is the field-metric
\beq \si_{ij} = \left(\begin{array}{cc}1 & 1 \\ -1 & 0\end{array}\right) ~~,\eeq
\noindent and the potential is given by 
\beqa && U(\varphi^1,\varphi^2) =  {1 \over 4} \exp \left( -{2 \sqrt{3}\over 3} \varphi^1 \right) \times \\ \nonumber &&  \left[\varphi^2 - 
{f_1(\varphi^2 )\over 2\ka }  \exp \left( -{2 \sqrt{3}\over 3} \varphi^1 \right)  \right]~~, \eeqa
\noindent Notice that any use of the metric in this Lagrangian density (contractions in kinetic terms, {\it etc}.) is done not with $\tilde{g}_\mn$, but with the physical metric $g_\mn = \exp[-(2/\sqrt{3})\varphi^1] \tilde{g}_\mn$.

\section{Lagrangian density of a perfect fluid}

The argument of the preceding section serves to show another striking feature of non-minimally coupled models: the Lagrangian density of matter appears explicitly in the field equations. In particular, one focuses on the case of a perfect fluid, given its ubiquity as a useful description for standard matter. Its energy-momentum tensor is the familiar form
\beq \label{emtensor} T_\mn = (\rho + p)u_\mu u_\nu + p g_\mn ~~,\eeq
\noindent where $\rho$ is the energy density, $p$ the pressure and $u_\mu$ the four-velocity (with $u_\mu u^\mu = -1$); its trace is $T = 3p -\rho$. The associated Lagrangian density is less discussed, precisely because it is absent from the field equations of GR: in applications of GR, the choice of Lagrangian density is mostly irrelevant, while it is of the utmost importance when considering a non-minimal coupling \citep{fluid}. 

The identification $\cl = p$ was first advanced in \citep{Seliger}, with a relativistic generalization in \citep{Schutz}. Much later, \citep{Brown} showed that this choice is equivalent to $\cl = -\rho$, complemented by a suitable set of thermodynamical potentials and Lagrangian multipliers $\phi^\mu = \varphi_{,\mu}  +s\theta_{,\mu}+\beta_A \alpha^{A}_{,\mu}$; these enable the relativistic thermodynamical relations via a current term $J_\mu \phi^\mu$ in the action, where $J_\mu$ is the vector density, {\it i.e.} the flux vector of the particle number density. In \citep{HE}, an isentropic perfect fluid is described via $\cl = -\rho$.

The equivalence between occurs on-shell, by substituting the field equations derived from the matter action
\beq  \label{actionfluid} S_m = \int\left( - \sqrt{-g} \rho + J_\mu \phi^\mu \right) d^4x~~, \eeq
\noindent back into the action functional and reads the resulting on-shell Lagrangian density $\cl_1 = p$. Similarly, the action may be rewritten so that the on-shell Lagrangian density reads $\cl_2 = na$, where $n = |J|/\sqrt{-g}$ is the particle number density and $a(n,T) = \rho(n)/n - sT$, where $s$ is the entropy per particle and $T$ is the temperature. Lastly, one may remove the current term $J_\mu \phi^\mu/\sqrt{-g} $ from \eq{actionfluid}, thus obtaining the on-shell Lagrangian density $\cl_3 = -\rho$ through the addition of adequate surface terms.

Indeed, \citep{Seliger} did not argue that $\cl_1 = p$ was the bare (as opposed to on-shell) Lagrangian density for a perfect fluid, but resorted to a much more evolved action functional: this simple identification with the pressure also arose only after the resulting field equations where substituted into it.

With the above in mind, one now ascertains how should this procedure be generalized to a non-minimally coupled scenario. This should affect the terms in \eq{actionfluid} that are minimally coupled, so that the matter action becomes
\beq
  S'_m=\int \left(- \sqrt{-g} f_2(R) \rho + J^\mu \phi_\mu \right) d^4 x~~,
  \label{modified fluid-action}
\eeq
\noindent while the current term remains uncoupled (aside from the use of the metric to contract indexes). Varying the action with respect to the potentials included in $\phi_\mu$, one obtains
\beqa &&- f_2 \mu U_\mu = \phi_\mu \,, \\ \nonumber &&T = {1 \over n}{\partial \rho \over \partial s}\Bigg|_n = {1 \over f_2(R)} \th_{,\mu}U^\mu \,, \eeqa
\noindent where $\mu$ is the chemical potential and $\th$ is a scalar field (included in $\phi_\mu$) whose equation of motion imposes the entropy exchange constraint $(s J^\mu)_{,\mu} = 0$. Thus, the non-minimal coupling of curvature to matter is reflected in both the velocity and the temperature identification.

By substituting the modified equations of motion into action (\ref{modified fluid-action}), an on-shell Lagrangian density $\cl_1 = p$ may be read, as in GR. The addition of surface integrals also yields the discussed $\cl_2 = -\rho$ and $\cl_3 = -na$.

Albeit the action may adopt distinct on-shell forms, this does not translate into a direct equivalence between them: only the original bare Lagrangian density $\cl_0$ should be inserted into the field equations \eq{field0}. However, this bare $\cl_0$ should not appear into the non-conservation law: indeed, when deriving \eq{cov}, the current term (which is not coupled with the metric) is dropped and one is indeed left with $\cl_2 = -\rho$.

Notwithstanding the above, one can adopt a simpler stance regarding the choice of the Lagrangian density $\cl = -\rho$ instead of $\cl = p$: if a dust distribution is to be considered --- that is, a perfect fluid with negligible pressure and corresponding energy-momentum tensor $T_\mn = \rho g_\mn $ ---, it appears unnatural to take a vanishing quantity as Lagrangian density.

Ending this section, one remarks that if a matter form is actually described by two independent bare Lagrangian densities leading to different dynamical behaviour of \eq{field0} and (\ref{cov}), then only confrontation with observation can ascertain which is the correct description.

\section{Dark matter mimicking}

One now reviews a mechanism that mimics the missing dark matter revealed through the flattening of galaxy rotation curves \citep{mimic}. In order to isolate the effect of the non-minimal coupling, one sets $f_1(R) = 1$ and assumes a power-law
\beq f_2(R) = 1 + \left({R \over R_n}\right)^n~~, \eeq
\noindent Since the dark matter contribution is dominant at large distances {\it i.e.} low curvatures, a negative exponent $n$ is expected. 

Inserting this into the modified field \eq{field0}, together with the Lagrangian density $\cl= -\rho$ for a dust distribution with $p=0$, yields
\beqa \label{field0mimic} && \left[ 1 - n \left( {R \over R_n} \right)^n  { \rho \over \ka R} \right] R_\mn - {1 \over 2} R g_\mn =  \\ \nonumber && \left[ 1 + \left( {R \over R_n} \right)^n \right] {\rho \over 2\ka} U_\mu U_\nu - n \De_\mn \left[ \left( {R \over R_n} \right)^n {\rho \over \ka R} \right] ~~. \eeqa
\noindent At large distances, normal matter is subdominant and the trace \eq{fieldtrace} reads
\beq \label{fieldtracemimic2} R  = (1-2n) \left( {R \over R_n} \right)^n  {\rho \over 2\ka }  -3n\square \left[ \left( {R \over R_n} \right)^n  {\rho \over \ka R} \right] ~~. \eeq
\noindent Inspection shows that an exact solution is obtained if the last term vanishes, 
\beq \label{exact} R = R_n \left[(1-2n) {\rho \over \rho_n}\right]^{1/(1-n)}~~.  \eeq
\noindent defining the characteristic density $\rho_n = 2 \ka R_n$.

A more evolved study of \eq{fieldtracemimic2} shows that the most general solution oscillates around the one above: this makes the gradient term actually dominate \eq{fieldtracemimic2}, and also allows for a perturbative $f_2(R) \ll 1$. As numerical results show, one may disregard these oscillations and simply consider \eq{exact} \citep{mimic}. A perturbative non-minimal coupling is paramount: it makes the mimicking mechanism satisfy both that the weak, strong, null and dominant energy conditions; grants immunity against Dolgov-Kawasaki instabilities \citep{DK}; and makes the extra force arising from \eq{cov} very small.

Instead of solving \eq{field0mimic}, obtaining a modified gravitational potential and then reading the mimicked dark matter contribution, one interprets the additional curvature obtained from the non-minimal coupling as due to the latter's density profile,
\beq \rho_{dm} \equiv 2\ka R = \rho_n \left[(1-2n) {\rho \over \rho_n}\right]^{1/(1-n)} ~~. \eeq
\noindent Thus, one obtains a direct translation between a visible matter profile $\rho$ and the mimicked dark matter distribution $\rho_{dm}$, with $n$ related to the large radius behaviour of both contributions.

The mimicked dark matter may be further characterized by inserting \eq{exact} into \eq{field0mimic}, and reading the obtained terms as due to a corresponding energy-momentum tensor: this allows one to obtain the EOS parameter for dark matter, $\om = p_{dm} / \rho_{dm}= n /( 1-n)$: for a negative exponent $n$, one has $\om < 0 $, hinting at at possible cosmological role in enabling an accelerated expansion of the Universe, as will be discussed in the following section.

\begin{figure} 

\epsfxsize=\columnwidth \epsffile{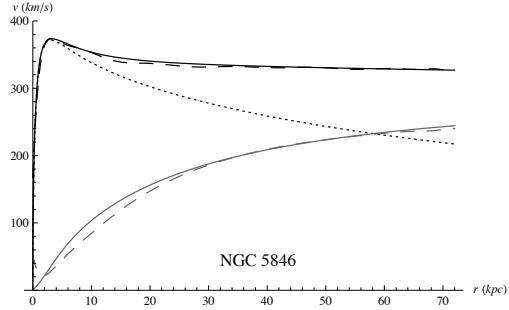}
\caption{Observed rotation curve (dashed full), decomposed into visible (dotted) and dark matter (dashed grey) contributions \citep{kronawitter}, with mimicked dark matter profile (full grey) and resulting full rotation curve (full).}
\label{rotation}

\end{figure}

In order to fit the galaxy rotation curves of several galaxies (NGC 2434, 5846, 6703, 7145, 7192, 7507 and 7626, which are almost spherical type E0 galaxies with well determined rotation curves \citep{kronawitter}), one resorts to the Hernquist profile for visible matter \citep{Hernquist}, which behaves as
\beq \label{Hernquistprofile} \rho(r) \sim {a \over r} \left(1+{r \over a} \right)^{-3}~~ , \eeq	
\noindent and the Navarro-Frenk-White \citep{NFW} and isothermal sphere profiles for dark matter: the latter enables a perfectly flat rotation curve, while the former is favored by numerical simulations:
\beq \label{dmprofiles} \rho_{IS}(r) \sim r^{-2}~~~~,~~~~\rho_{NFW}(r) \sim {a \over r} \left(1 + {r \over a} \right)^{-2} ~~. \eeq
\noindent with $a$ signalling the transition between inner and outer slope profiles. Given the outer slope behaviours $\rho(r) \sim r^{-4}$, $\rho_{IS} (r) \sim r^{-2}$ and $\rho_{NFW}(r) \sim r^{-3}$, the relation $\rho_{dm}(r) \propto \rho(r)^{1/(1-n)}$ yields the exponents $n_{IS} = -1$ and $n_{NFW}= -1/3$ that translate the Hernquist profile into the two above. Thus, one is led to consider the composite non-minimal coupling
\beq f_2(R) = 1 + {R_{-1} \over R} + \sqrt[3]{R_{-1/3} \over R}~~. \eeq
The values of the model parameters $R_{-1}$ and $R_{-1/3}$ are obtained from fitting the result of the numerical integration of \eq{field0mimic} to the available galaxy rotation curves: by doing so, one concludes that $R_{-1} \sim 1/(16.8~Gpc)^2$ and $R_{-1/3} \sim 1/(1.45 \times 10^6 ~Gpc)^2$. Fig. \ref{rotation} shows the fit to NGC 5846, one of the galaxies addressed in \citep{mimic}.

Allowing for individual fits of these parameters to each of the considered galaxies shows that there is some dispersion around these orders of magnitude, which could be due to deviations from sphericity, poor choice of visible or dark matter density profiles, unaccounted effect of a non-trivial $f_1(R) \neq R$ term or the non-minimal coupling with the electromagnetic sector (thus enabling a dependence on luminosity) {\it etc.}. Nevertheless, the quality of the obtained fits and the elegance of the mimicking mechanism support its ability to account for the missing dark matter puzzle.

\section{Accelerated expansion of the Universe}

This section discusses the possible use of a non-minimal coupling to describe the current phase of accelerated expansion of the Universe \citep{accexp}. One focuses on a constant deceleration parameter $q \equiv -\ddot{a}a / (\dot{a})^2 = 1/\be - 1$, which translates into a power-law expansion with the scale factor evolving as $a(t) = a_0 (t/t_0)^\be$, with $t_0 = 13.73 ~Gy$.

This hints at the use of a power-law for the non-minimal coupling also, as considered in the previous section. Thus, one takes $f_2(R) = 1 + (R/R_n)^n$, with the exponent $n$ assumed to be negative --- so that the accelerated expansion phase appears at late times, when $R \ll R_n$.

This accelerated expansion is obtained in a rather straightforward fashion: assuming a flat Friedmann-Robertson-Walker metric with line element $ ds^2 = - dt^2 + a^2dV^2 $ and that matter is described by a perfect fluid with density $\rho$ and pressure $p$, one has $ T_{00} = \rho $ and $T_{rr} = a^2 p$. One then uses $\cl= -\rho$ and \eq{cov} to ascertain that energy is covariantly conserved in a cosmological context,
\beqa && \nabla_\mu T^{\mu 0} = {F_2 \over f_2}\left(g^{0 0} \cl-T^{0 0}\right)\dot{R} = 0 \rightarrow \\ \nonumber && \dot{\rho} +3H\rho = 0 \rightarrow \rho(t) = \rho_0 \left({t_0\over t}\right)^{3\be}~~. \label{covcosmo} \eeqa
\noindent where $\rho_0 = \Om_m \rho_{crit}$, with $\Om_m \sim 0.3$ the relative matter density and $\rho_{crit}  \sim 10^{-26}~kg/m^3 $ the critical density \citep{WMAP7}.

One uses this result together with \eq{field0} to compute the modified Friedmann equation
\beq \label{modFriedmann} H^2 = {1 \over 6\ka} (\rho + p + \rho_c + p_c)~~, \eeq
\noindent where the additional density $\rho_c$ and pressure $p_c$ terms are introduced,
\beqa \rho_c & =& -6 \rho_0 \be {  1-2\be + n(5 \be + 2n -3)  \over  \left({t\over t_0}\right)^{3\be} \left({t \over t_2}\right)^{2n} \left[6\be (2\be-1)\right]^{1-n} } ~~, \\ \nonumber p_c &=& -2 \rho_0 n { 2+4n^2 -\be (2+3\be) +n(8\be -6) \over  \left({t\over t_0}\right)^{3\be} \left({t \over t_2}\right)^{2n} \left[6\be (2\be-1)\right]^{1-n} }~~.   \eeqa
\noindent defining $t_n \equiv R_n^{-1/2}$ and taking the weak regime $F_2 \rho \ll \ka$ \citep{accexp}.

\begin{figure} 

\epsfxsize=\columnwidth \epsffile{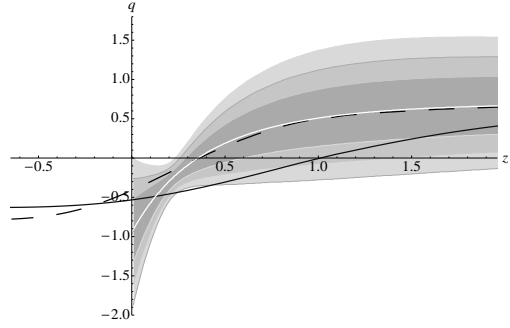}
\caption{Evolution of the deceleration parameter $q(z)$ for $n=-4$, $t_2=t_0/4$ (full) and $n=-10$, $t_2=t_0/2$ (dashed); $1\si$, $2\si$ and $3\si$ allowed regions are shaded, white line gives best fit \citep{gong}.}
\label{qzgraph}

\end{figure}

Given the Hubble parameter $H(t) \equiv \dot{a}/a = \be/t$, the {\it l.h.s.} of the Friedmann \eq{modFriedmann} falls as $t^{-2}$, so that comparing with the above gives a relation between the exponents $\be$ and $n$:
\beq 3\be + 2n = 2 \rightarrow \be = {2 \over 3}(1-n)~~. \eeq
\noindent Thus, one concludes that any negative exponent $n$ will yield an accelerating Universe with $\be$, $q > 0$. One may also compute the deceleration parameter and equation of state parameter for the non-minimally coupled contribution,
\beq q = - 1 + {3 \over 2(1-n)} ~~~~,~~~~\om = {n \over 1 - n}~~. \label{eqqz} \eeq
\noindent The latter is the same form found in the previous section. By fitting the numerical solution of \eq{modFriedmann} to the evolution profile of $q(z)$ \citep{gong}, one obtains the best fit values $n = -10$ and $t_{-10} = t_0/2 $, depicted in Fig. \ref{qzgraph}.

This value for the exponent $n$ is very distinct from the $n_{IS} = -1$ and $n_{NFW} = -1/3$ scenarios considered in the previous section. These have no cosmological impact on the current scenario, given the broad difference between the relevant timescale $t_0$ and those obtained from $t_{-1} \equiv R_{-1}^{-1/2}$ and $t_{-1/3} \equiv R_{-1/3}^{-1/2}$. Its value $n = -10$ may also be regarded as somewhat unnatural, but is highly dependent on the $q(z)$ profile: Fig. \ref{qzgraph} shows that $n = -4$, $t_2 = t_0/4$ also yields a curve within the allowed $3\si$ region. Despite this caveat, one concludes that a non-minimal coupling may be used to describe the accelerated expansion of the Universe.

\section{Matter solution with constant curvature}

The main result of the previous section is embodied by \eq{eqqz}, which directly translates the exponent $n$ of a power-law coupling between curvature and matter to the deceleration parameter $q$. However, it is trivial to check that one cannot produce a de Sitter Universe, characterized by $q = -1$, as this would require $n \rightarrow \infty$ --- so that a Cosmological Constant (CC) $\La$ with EOS $p_{\La} = -\rho_\La$ cannot be recovered. One could argue that this hints at the need to consider a different form for $f_2(R)$ --- perhaps an exponential, given the behaviour of the scale factor $a(t) = a_0 \exp (H_0t)$.

However, as was shown in \citep{accexp}, it turns out that a de Sitter phase is impossible to obtain in a cosmological context, fundamentally because the curvature is given by the CC, $R = 4\La$, and is thus constant: since the density still evolves as $\rho(t) \propto a^{-3} \sim \exp(-3H_0t)$, the terms arising from the non-minimal coupling in the modified Friedmann cannot be interpreted as an additional density and pressure due to a CC, $p_{CC} = -\rho_{CC}$. One may simplistically state that, since the $f_2(R) \cl$ term in \eq{model} cannot be made constant, a CC is unattainable.

With the above in mind, this section briefly discusses the possibility that a CC is not directly generated in a cosmological context, but arises due to an averaging of several constant curvature ``bubbles'' \citep{localCC}. This does not aim at solving the Cosmological Constant (CC) problem: the $\sim 10^{120}$ discrepancy between its expected value and that needed to account for the current accelerated expansion, but draws a mechanism to produce the latter \citep[for a review, see][]{OB2009}.

Notice that a constant curvature $R = R_0$ implies that all terms involving $f_2$ and its derivatives become constants; thus, the actual form of $f_2$ is shown to be only constrained by $F_2(4\La)\La \ll  f_2(4\La)$. In particular, the linear or power-law non-minimal couplings considered in this work are compatible with this relation.

\begin{figure} 

\epsfxsize=\columnwidth \epsffile{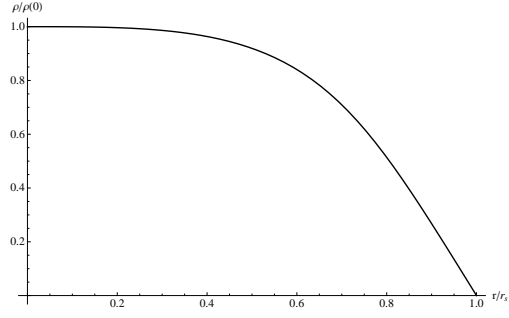}
\caption{Rescaled density profile arising from $R_0 = 4\La$.}
\label{graphrholocalCC}

\end{figure}

One uses a static Birkhoff metric with line element $ds^2 = - e^{2\nu(r)} dt^2 + e^{2\si(r)} dr^2 + d\Om^2$; matter is assumed to be described by a perfect fluid with constant EOS parameter $\om = p/\rho$. The Einstein field \eq{field0} with $f_1(R) =R $ thus leads to the differential equation,
\beqa  \label{difeqlocalCC} && 0 = \rho'' - {2 \over r} \rho' + {1+\om \over 2(KR_0r)^2}{f_2 \over F_2} \rho (\rho')^2 - \\ \nonumber && - {1\over 3r^3} \left({ 1 + \om \over 2(KR_0)^2}\right)^2 \left({f_2 \over F_2}\right)^2  \rho^2 (\rho')^3 + \\ \nonumber && + {1 \over 6(KR_0)^2r} (\rho')^3 \left( {2 \over r^2} - R_0 + {1 + \om \over 2}{f_2 \over F_2} \right) ~~,\eeqa
\noindent where $K$ is an integration constant. Imposing continuity with a  Schwarzschild-de Sitter metric at the boundary of the matter distribution $r = r_s$, one obtains
\beq K = {\rho'(r_s) \over R_0} \sqrt{{1  \over 3r_s^2} -\La }~~. \eeq
The boundary radius $r_s$ is related to the total mass of the matter distribution via
\beq r_s \sim \sqrt{\rho_H \over \rho_c} 11.4~\mpc ~~, \eeq
\noindent where $\rho_H = 1~ \mathrm{H ~atom / cm^3} = 1.660 \times 10^{-21} ~ kg / m^3$.
The (rescaled) numerical solution to \eq{difeqlocalCC} when $R_0 = 4\La$ ({\it i.e.} the local curvature yields a CC) is shown in Fig. \ref{graphrholocalCC}. The typical distance between galaxies is of the order of  $r_s \sim 1 \mpc$ ({\it e.g.} Andromeda is $778~\mathrm{Kpc}$ away), so that $ \rho_c \approx 130 ~\mathrm{H ~atom / cm^3} $ --- close to the lower bound of $H II$ star-forming regions. The interstellar medium attains densities up to $10^4 ~\mathrm{H ~atom / cm^3}$, allowing for a radius of the constant curvature bubble down to $\sim 100~\mathrm{Kpc}$. Thus, one concludes that it is possible to obtain a smooth constant curvature matter solution $R_0 = 4\La$ compatible with the expected values for realistic density profiles.

\subsection{Averaging mechanism}

Having obtained a feasible description of one constant curvature ``bubble'', one now briefly discusses how several of these may average and produce a CC at cosmological scales. One first argues that a universal value of $R_0=4\La$ is to be expected in all local ``bubbles'', as it should be derived from the non-minimal coupling $f_2(R)$ itself: a full account of the dynamics of the model, obtained by allowing the curvature to vary, should lead to a relaxation towards a constant value $R_0$ arising from the modified field \eq{field0}, if the composite effect of a non-trivial curvature term $f_1(R)$ and coupling $f_2(R)$ exhibits a suitable minimum, thus acting as an attractive potential. 

This minimum $R_0(\rho)$ would naturally depend on the underlying density, suggesting a similarity with the so-called chameleon model \citep{chameleon} --- indeed the full Lagrangian $f_1(R) - f_2(R) \rho$ may be suggestively compared with the usual chameleon effective potential $V_{eff}(\phi) = V(\phi) + \rho \exp(\phi/M)$. At early times, local density and curvature solutions could be widely different from each other and from the constant curvature and specific matter distribution assumed here: however, extensive neighbour to neighbour interaction would favour similar density profiles, yielding a universal minimum $R_0$. This would lead to a progressive decrease of the deceleration parameter towards $q = -1$, signaling the onset of a CC \citep{gong}.

\section{Post-inflationary reheating}

In this section one discusses the possibility of using a non-minimal coupling to drive the reheating of the Universe after the inflationary stage occurring in its early times \citep{preheatingf2}. One does not attempt to drive the dynamics of inflation through the effect of $f_2(R)$, but instead assumes the well-known Starobinsky inflation model \citep{Staro}--- which resorts to a quadratic curvature term
\beq f_1(R) = R + {R^2 \over 6M^2}~~,\label{staroform}\eeq
\noindent with $M \sim 3 \times 10^{-6}M_P$.

A non-minimal coupling is fundamental in the so-called preheating mechanism: the reheating of the ultracold post-inflationary Universe due to the explosive production of particles, occurring when the dynamics of a quantum scalar field $\chi$ endowed with a variable mass term of the form $m^2_{eff} = m^2 + \xi R$ experience parametric resonance \citep{preheating}. More evolved couplings also lead to preheating, as found in Refs. \citep{mimoso,kinetic}.

This hints that the non-minimally coupled action (\ref{model}) may generalize the preheating scenario. Given the form of the variable mass term $m_{eff}$, one assumes a linear coupling 
\beq f_2(R) = 1 + 2\xi {R \over M^2}~~. \eeq
Since the curvature is  coupled to matter and radiation, besides the scalar field $\chi$, one must ensure that the cosmological dynamics are driven by the effect of the quadratic curvature term \eq{staroform} alone --- {\it i.e.} that the non-minimal coupling only intervenes during preheating, not before. As shown in \citep{preheatingf2}, this implies a perturbative regime $f_2(R) \sim 1$ and that $1 < \xi  < 10^4$, compatible with the weak bound $\xi \ll 10^{78}$ obtained from considerations on solar hydrostatic equilibrium \citep{hydro}.

Decomposing the scalar field $\chi$ into its Fourier modes $\chi_k$, one finds that they follow the differential equation
\beqa \label{equationfourier} X''_k + \bigg[\left({ 2 k \over a M}\right)^2 + \left({2m \over M}\right)^2 - 3 {H' \over M} - 9 {H^2 \over M^2} + \\ \nonumber {\xi \over M^2} \left( \xi {R'^2 \over M^2} - 6 {H R' \over M}- R'' \right) \bigg] X_k = 0~~. \label{hilleq} \eeqa
\noindent using the redefinition $X_k \equiv a^{3/2} f_2^{1/2} \chi_k \sim a^{3/2} \chi_k$ and the new variable $2z = M(t-t_o) \pm \pi/2$ (depending on the sign of $\xi$).

After slow-roll, the Hubble parameter $H(t)$ experiences an oscillatory phase, so that $R \sim (2M^2 / z) \cos(2z) $: the $R''$ term dominates and one rewrites the above as a Mathieu equation,
\beq X''_k + \left[A_k - 2q \cos(2z) \right] X_k = 0~~, \label{mathieu} \eeq
\noindent with
\beq A_k = \left({2k \over a M}\right)^2 + \left(2 m \over M\right)^2 ~~~~, ~~~~ q = {4 \xi \over z}~~.\eeq
\eq{mathieu} is the exact same form found in usual preheating \citep{preheating}: the quantum field $\chi$ experiences parametric resonance as the scale factor $a(t)$ increases, with massless particles produced for a coupling parameter as low as $\xi \gtrsim 3$, while massive particles require $\xi \gtrsim 10$ --- well within the range $ 1 < \xi < 10^4$ discussed above. Thus, one finds that, as expected, a universal non-minimal coupling may successfully drive the reheating of the post-inflationary Universe.

\section{Conclusions and Outlook}

A non-minimal coupling between geometry and matter covers a broad spectrum of applications in an elegant, natural way. Hopefully, this hints at its relevance in the formulation of a fundamental quantum theory of gravity: future advances on the theoretical front should help clarify its nature.

The distinct expressions for the non-minimal coupling here considered may be regarded as an approximation to a more evolved forms for $f_2(R)$, each valid in a particular regime: early {\it vs.} late time, central {\it vs.} long range, {it etc}.. The latter could perhaps be written as a Laurent series,
\beq f_2(R) =  \sum_{n = - \infty}^{\infty} \left({R \over R_n}\right)^n ~~, \label{Laurent} \eeq
\noindent so that one could chart yet unprobed terms of the above series by assessing the dynamics of other phenomena and environments, where distinct curvatures and densities are at play.
?\begin{acknowledgements}
This work was developed in the context of the GREAT-ESF Workshop ``QSO Astrophysics, Fundamental physics, and Astrometric Cosmology in the Gaia era'', 6-9 June 2011, Universidade do Porto, Portugal. I wish to thank the organization, specially Sonia Anton and Mariateresa Crosta, for  the invitation and hospitality.
\end{acknowledgements}
\bibliographystyle{aa}

\end{document}